\documentclass[aps,12pt]{revtex4-1}
\usepackage{amsmath,amsfonts}
\usepackage{graphicx}
\newcommand\BZ{{\mathbb Z}}
\newcommand\tr{\hbox{Tr}}
\newcommand\vev[1]{\langle #1 \rangle}
\newcommand\ket[1]{| #1\rangle}
\newcommand\bra[1]{\langle #1|}
\newcommand\CO{{\cal O}}

\usepackage[all]{xy}
\begin{document}
\title{A toy model for  time evolving QFT on a lattice with controllable chaos}

\author{David Berenstein}
\affiliation {Department of Physics, University of California at Santa Barbara, CA 93106}

\begin{abstract} 
A class of  models with a dynamics  of generalized quantum cat maps on a product of quantum tori is described.  These tori are defined by an algebra of clock-shift matrices of dimension $N$.
The dynamics is such that the Lyapunov exponents can be computed analytically at large $N$.
Some of these systems can be thought of as a toy model for quantum fields on a lattice under a time evolution with nearest neighbor interactions, resembling a quantum version of a cellular automaton.
The dynamics of entangling is studied for initial product states. Some of these entangle at rates determined by Lyapunov exponents of the system at large $N$ when the initial states are gaussian.
For other classes of states, entanglement between two regions can be computed analytically: it is found that entanglement rates are controlled by $\log(N)$.  
Some of these setups can be realized on quantum computers with CNOT quantum gates. This is analyzed in detail where we find that the dynamics has a self-similar behavior and various peculiar behaviors. This dynamics can be interpreted in a particular basis as 
a machine that broadcasts classical messages in one direction and that produces over time a generalized GHZ state with the receiving region once we consider superpositions of such messages. 
With the appropriate choice of quantum vacuum on the receiving end of the system one can stop the message from leaving the broadcast area. 
\end{abstract}
\maketitle

\section{Introduction}

Recent advances in understanding the dynamics of holographic field theories have found interesting connections between properties of black holes and quantum chaos \cite{Shenker:2013pqa}. One of the main observations in this development by Shenker and Stanford   is that the butterfly effect, expressed by having non trivial Lyapunov exponents for a dynamical system,  is encoded in the dynamics of the near horizon geometry of the black hole. In particular the Lyapunov exponent that is computed this way is the temperature of the black hole (to get the units to work one needs to use a normalized $k_BT/\hbar$,where $k_B$ is the Boltzmann constant). A bound on this exponent can be proved analytically for holographic-like systems using analyticity of out of time ordered correlations functions (OTOC) \cite{Maldacena:2015waa}. This bound is restricted to near-equilibrium conditions and depends on an ensemble averaging at finite temperature. At least one calculable model, the SYK model, is known to satisfy this bound \cite{Maldacena:2016hyu}.

Lyapunov exponents also appear in the dynamics of entanglement \cite{Zurek:1994wd}. It has been argued that if one selects $k$ degrees of freedom in a semiclassical quantum chaotic system, then these will generically entangle at a rate that is controlled by the end sum of the $2k$ largest Lyapunov exponents \cite{Asplund:2015osa}.  What not clear is for what classes of states such characterization is correct. A proof for linear dynamics can be found in \cite{Bianchi:2017kgb}, which is applicable to Gaussian states. The author is currently writing a proof of a bound for general non-linear  dynamical systems related to smooth symplectomorphisms of a finite volume phase space in the $\hbar \to 0$ limit \cite{BerensteinKS},  for initial conditions that are semiclassical (of low uncertainty on the manifold). Numerical simulations for a simple example can be found in \cite{Casati}. 

Such an entanglement rate based on Lyapunov exponents should be  approximately valid until the entropy of individual degrees of freedom saturates. In previous work of the author with A. Garcia Garcia \cite{Berenstein:2015yxu}
it was argued that this provides a mechanism for having general bounds on Lyapunov exponents at finite $\hbar$.
The saturation of the entropy is expected to  occur on a time scale which is of the order of the Ehrenfest time, which is logarithmic in $\hbar$. This is argued in the semiclassical approximation, but it is not clear that it applies to other states.  Part of the aim of this paper is to shed some light on this issue in a class of controllable examples.

Entanglement production in quantum field theories is more poorly understood, as it is difficult to find oneself in the semiclassical regime. In holographic theories, entanglement production can be understood for some class of states where gravity is classical in terms of the Ryu-Takayanagi formula \cite{Ryu:2006bv} and the HRT proposal \cite{Hubeny:2007xt}, whereas 
for conformal field theories in two dimensions one gets some intuition from the Cardy-Calabrese result \cite{Calabrese:2004eu}, which depends on a particular quench from a very particular class of initial states.

One of the main problems with chaotic dynamical systems is that in general one has to compute their Lyapunov exponents numerically, and under most circumstances the quantum simulations of the dynamics are forbidding due to the sign problem and the exponential growth of the size of the Hilbert space in terms of the number of sites on a lattice. Holographic theories need of order $N^2$ degrees of freedom, and although some progress has been made on computing the Lyapunov exponents  in the classical theory \cite{Gur-Ari:2015rcq},
very little is understood about direct quantum computations of entanglement production.

Given these circumstances, it is desirable to have models of quantum field theories where one can compute analytically both the Lyapunov exponents of the theory and the entanglement entropy evolution between regions (at the very least for some simple setups). 

This article provides such an example. The example describes a quantum system with a discrete time evolution. The system is a generalization of the classical  Arnold cat map on a torus to a generalized version of the dynamics to many such tori. 
It is possible to set up 
examples where the 
dynamics on each iteration is of nearest neighbor type and that it mimics a lattice quantum field theory. In a particular basis, the quantum evolution examples we discuss can also be understood as a classical evolution on a lattice with discrete values at each site, mimicking a reversible cellular automaton. 

The system will be such that the Lyapunov exponents can be computed analytically by diagonalizing a single matrix. 
Because the dynamics is discrete, it is hard to argue that it can be analyzed at a fixed temperature: the evolution is not due to time evolution with a Hamiltonian, and therefore the usual analytic continuation to imaginary time is not possible. 
As such, these models have very little to say about the (temperature dependent) quantum bound on chaos \cite{Maldacena:2016hyu}.
On the other hand, the entanglement dynamics of the system can be analyzed in detail in a particular class of examples, and some analytic results are possible. Here we find that the entanglement entropy production between regions is bounded above  by the (logarithm of the ) size of the Hilbert space of a single site. This is independent of Lyapunov exponents. In some other (semiclassical) regimes entanglement is controlled by the Lyapunov exponents of the system. This way we answer partially the problem of when the Lyapunov exponents bound the entropy growth: they do so for semiclassical states, but not for generic states. 

The article is organized as follows. In section \ref{sec:quantumcatmaps}  quantum cat maps on a quantum torus re reviewed. In section \ref{sec:gqcm} I show a generalization of quantum cat maps on a lattice that is very easy to analyze. Apart from the classical dynamics, it has an additional parameter $N$ that is the dimension of the Hilbert space s at one site. The evolution is described as an automorphism of the quantum torus algebra in a manner that is essentially $N$ independent. In particular, the Lyapunov exponents can be easily computed from the automrphism. In section \ref{sec:twosite} the simplest model with two sites is analyzed and entanglement entropy is computed as a function of time for a variety of states. For two types of basis states, there is no entanglement entropy generation, but results for more general classes of states are also analyzed. It is shown that one can have situations where the maximal entropy ($log(N)$)  is essentially attained in one shot, and for initial semiclassical states the entropy production is controlled by the sum of the two largest Lyapunov exponents. In section \ref{sec:CNOTqft} a particularly simple lattice field theory is analyzed. If the  local site Hilbert space has dimension $N=2$ the evolution is generated by CNOT quantum gates. These systems mimic cellular automata in a particular basis of states. The dynamics exhibits interesting entanglement behavior that can be controlled for a class of states and the entanglement  entropy production rate is bounded above by $\log(N)$. The dynamics in a  simple basis acts as a classical machine that can send (encoded) messages of any length in one direction. In a dual (Fourier) basis, it acts as a machine that sends messages in the opposite direction. Utilizing these insights one can show that depending on the choice of {\em vacuum} on the receiver side, one can stop the entanglement entropy between both sides from being generated, even though one would expect that classically the messages will always be delivered on the other side. I conclude in section \ref{sec:conclusion}.

\section{Cat maps and their quantization}\label{sec:quantumcatmaps}

Consider a 2D-torus $T^2\simeq S^1\times S^1$ with two angle coordinates $x_1,x_2 \in [0,2\pi]$ as a phase space. The cat maps are linear maps that preserve the area (and hence the Poisson brackets). They are characterized by
\begin{equation}
\begin{pmatrix}
x'_1\\
x'_2
\end{pmatrix}=\begin{pmatrix} a & b\\
c & d\end{pmatrix}\begin{pmatrix}
x_1\\
x_2
\end{pmatrix}= M \vec x
\end{equation}
and the periodicity requires that $a,b,c,d $ are all integers and area preserving condition requires that $ad-bc=1$, so the set of such maps is $SL(2,\BZ)$. The properties of the map essentially only depend on the trace $a+d$. The famous Arnold cat map is the one with $a+d=3$. The logarithm of the eigenvalues of the matrix $M$ are the Lyapunov exponents of the system, when one iterates the dynamics. Their classical dynamics is discussed in \cite{Zaslavsky} for example. Their realization as quantum system as was described in \cite{HANNAY1980267} (see also \cite{Keating}).

The eigenvalues of $M$ are the two solutions to the quadratic equation
\begin{equation}
\gamma^2-(a+d) \gamma +1=0
\end{equation}

If one of the eigenvalues is in norm bigger than one, then the system is chaotic. This condition requires that $|a+d|> 2$. These are very well known and studied systems and they are a model of chaos, where the Lyapunov exponents can be computed easily and analytically.

To quantize, we can not use the variables $x_1, x_2$ directly because they are multivalued. Instead, one uses
\begin{eqnarray}
\hat P = \exp(i x_1)\\
\hat Q= \exp(i x_2)
\end{eqnarray}
These are the clock an shift matrices. From now on I will remove the hat that distinguishes an operator  and call the $P,Q$ themselves just matrices. They are defined as unitaries that satisfy the following algebra
\begin{eqnarray}
P^n=1\\
Q^n=1\\
PQ= \omega QP 
\end{eqnarray}
where $\omega^N=1$ is a primitive root of unity.  The effective $\hbar\propto [x_1,x_2] $ is $1/N$, and is deduced from the Baker-Campbell-Hausdorff formula as if the $x$ are  well defined single-valued variables. Consider that $\omega =\cos(2\pi/N)+i \sin( 2\pi/N)\simeq 1 +i \frac{2\pi i}N$ is an approximate solution
for the N-th root of unity. In this sense $2\pi i \hbar \simeq (\omega-1)$.
This can be intuited from
\begin{equation}
PQ-QP=(\omega-1) QP\simeq i \hbar \{Q,P\}
\end{equation}
where the curly brackets indicate the Poisson bracket.

The $P,Q$ matrices have a unique irreducible representation of dimension $N$.
The construction is as follows. Since $Q$ is unitary, its eigenvalues are complex unitaries and the eigenvectors are orthogonal. 
Using $Q^n=1$, one can check that the eigenvalues are $n$-th roots of unity. These can be labeled by $k$, where $\omega^k$ is the corresponding root of unity. Start with one such eigenvector (assume that it is orthonormal) and call it $\ket k$, which has eigenvalue $\omega^k$ for $Q$.
Clearly
\begin{equation}
P Q \ket k = \omega Q P \ket k = \omega^k P\ket k
\end{equation} 
So that 
\begin{equation}
P\ket k \propto \ket{k-1}
\end{equation}
And the constant of proportionality is an unitary number. This is deduced immediately from $P^\dagger P=1$. We can choose phases so that
\begin{equation}
P\ket k= \ket{k-1}
\end{equation}

Since $P^N=1$, it must be the case that $\ket{k-N}= \ket k$. Notice that this way the action of $P$ generates all possible roots of unity, each of them generated only once. The dimension of the representation is $N$.
The algebra generated by $P,Q$ is equivalent to the set of $N\times N$ complex matrices acting on this representation.

The set of matrices $F_{ k,\ell}=Q^k P^\ell$, with $k,\ell\in\{0, \dots, N-1\}$ are a complete set of matrices. They are orthogonal to each other under the trace norm, so that the following property holds
\begin{equation}
\tr (F_{k, \ell} F^\dagger_{k', \ell'} )= N \delta_{k, k'}\delta_{\ell,\ell'}\label{eq:orth}
\end{equation}

The quantized cat map (up to phases for the action on the algebra) is given by the automorphism of the algebra
\begin{eqnarray}
P'\simeq  Q^b P^a\\
Q' \simeq  Q^d P^c
\end{eqnarray}
I'm writing Q before P as a normal ordering choice.
The condition to be an automorphism is exactly that $ad-bc=1$. This means that to each element of $SL(2,\BZ)$ we can assign an automorphism of the  $P,Q$ algebra. Since this is an automorphism of the algebra of $N\times N$ matrices, and all such automorphisms are due to conjugation by an unitary, we find that the automorphism must be generated by a Unitary transformation acting on the vector space where the algebra is represented.

In that sense, it makes sense to give explicit realizations of the generators of $SL(2,\BZ)$. Once those are given, the rest follows. 

Recall that we have constructed the representation as follows
\begin{eqnarray}
Q&=&\begin{pmatrix}1 & & &\\
& \exp(2\pi i /N) & &\\
& & \ddots &\\
&&& \exp( 2\pi i (N-1)/N) \end{pmatrix} \nonumber\\
 P&=& \begin{pmatrix}0 &1 & &\\
&0  &1 &\\
& & \ddots &\ddots\\
1&&&0\end{pmatrix} 
\end{eqnarray}
The $T$ generator sends $Q\to Q, P\to QP$, corresponding to the matrix 
\begin{equation}
M= T=\begin{pmatrix}
1&1\\
0&1
\end{pmatrix}
\end{equation}
This leaves $Q$ invariant, so it should be realized by a diagonal unitary operator that commutes with it. One can check that with such an  ansatz $U\simeq \hbox{diag} (\exp i \theta_k)$
\begin{equation}
U P U^{-1} = Q P
\end{equation}
requires that 
\begin{equation}
\theta_k-\theta_{k+1}=\frac{2\pi i}{N} k
\end{equation}
Which is easily solvable (starting from $\theta_0=0$) for example.

Similarly, the $S$ transformation 
$$
M=S=\begin{pmatrix}0&1\\
-1&0\end{pmatrix}
$$
is described by a discrete Fourier transform. Since any other matrix belonging to $SL(2,\BZ)$ can be built from the generators, it is straightforward (but tedious) to build a Unitary that generates the wanted transformations up to  getting the phases correctly. To fix these phases (for example to move $Q$, from the right to the left in an expression), one conjugates by powers of $P,Q$
to reorder the monomials.

A rather interesting question is to understand the semiclassical limit $\hbar \to 0$ and how the physics depends on $\hbar$ when the dynamics associated to $M$ is fixed.
It is in this limit that one can compute accurately the quantum Lyapunov exponents of the system and relate them to quantum properties of the dynamics at finite time. We will do that now.

\subsection{Quantum Chaos and OTOC} 
 
I will now detail two points of view on how the Lyapunov exponents appear in real time correlation functions in quantum field theory and will them compute them in the case of the cat map dynamics.  The first point if view is in semiclassical physics, and the second point of view is statistical. They both amount to different 
ways of making a quantity with an indefinite sign positive.
 
 \subsubsection*{From Kubo's formula to Lyapunov exponents}
 
 Consider a system, with a time evolution dynamics that is perturbed by an (Hermitian) operator $\CO(t_0)$ at some fixed time $t_0$. The standard linear response theory (as in Kubo's formula) states that the future observable $A(t)$ (which is also assumed to be Hermitian) is modified by the following relation
 \begin{equation}
 \delta\vev{A(t)} = i \bra\psi [A(t),\CO(t_0)]\ket \psi
 \end{equation} 
 In the semiclassical limit $\hbar \to 0$ we should be able to replace the commutator by a Poisson bracket as follows \cite{larkin1969}
  \begin{equation}
 \delta\vev{A(t)} = - \hbar \bra\psi \{A(t),\CO(t_0)\}_{P.B.}\ket \psi
 \end{equation} 
 
This quantity is in general not positive definite and should average to zero on most systems. However, if the initial state is very classical and both $A(t)$ and $\CO(t_0)$ are sufficiently smooth classical function on phase space, we should be able to replace the right hand side by evaluating the Poisson bracket on the classical trajectory itself. What we get this way is 
  \begin{eqnarray}
 \delta\vev{A(t)} &= &- \hbar  \{A(t),\CO(t_0)\}_{P.B.} \nonumber \\
 &=& - \hbar \frac{\partial A(t)}{\partial \xi^J (t)} \frac{\partial \CO(t_0)}{\partial \xi^I(t_0)} \frac{\partial \xi^J(t)} { \partial \xi^{J'}(t_0)}  \omega^{J'I}\\
 &\propto& \frac{\partial \xi^J(t)} { \partial \xi^{J'}(t_0)}
 \end{eqnarray} 
 where $\omega^{J'I}$ is the symplectic form for a complete set of coordinates $\xi^I$. This expression is proportional to the Jacobian of the change of variables and therefore grows, in absolute value, like the exponential of the largest Lyapunov exponent of the system (see  \cite{Zaslavsky} for example).

 We get this way that 
 \begin{equation}
|  \delta\vev{A(t)} | \propto \exp(\kappa_+ (t-t_0))
 \end{equation}
 the response to a perturbation grows exponentially in time. If $A$ is bounded, the left hand side saturates and this indicates that the semiclassical approximation is breaking down. The time scale for this is of the order of the Ehrenfest time. 
 A slightly different way to use this information is via the generalized uncertainty relation \cite{Berenstein:2015yxu}, where
 \begin{equation}
 \Delta A(t) \Delta \CO(t_0) \geq \frac 12|\vev{[A(t),\CO(t_0)]}|_\psi \propto \left|\frac{\partial \xi^J(t)} { \partial \xi^{J'}(t_0)}\right|\simeq  \exp(\kappa_+ (t-t_0))
 \end{equation}
 so that the product of uncertainties is growing exponentially. If the state is fixed at time $t_0$, and the initial uncertainty is small (we have a semiclassical initial state), then over time the capacity to do detailed predictions on the observable $A(t)$ deteriorates exponentially like the leading Lyapunov exponent. 
This is  what is expected as a generic behavior. What we get is usually a  particular linear combination of elements of the Jacobian of the change of variables under time evolution. These linear combinations can be non generic for specific observables. What should be true is that we should be generically able to find pairs of observables whose growth of uncertainties grows exponentially. Again, saturation of the uncertainty indicates that we have hit the Ehrenfest time.

 Also notice that the response function involves always a set of operators that has the wrong time ordering in one of the terms of the commutator. 
 
 After this discussion, we can go ahead and evaluate the commutators of the basic (Unitary) operators $P(t)$, $Q(t)$, relative to $P(0), Q(0)$. These are not hermitian, but all we care about is the norm of the commutators at future time.
 For a transformation associated to the matrix 
 \begin{equation}
 M(t)= \begin{pmatrix} a&b\\
 c&d 
\end{pmatrix}= M_1^t
 \end{equation}
 we find that (up to a global phase for each of $P(t)$, $Q(t)$ that we have not determined yet)
   \begin{eqnarray}
 [P(t), P(0)] &=& (1-\omega^b)  Q^b P^{a+1} \label{eq:r1} \\ \ 
 [P(t), Q(0)] &=& (\omega^a-1) Q^{b+1} P^a\\ \
 [Q(t), P(0)] &=& (1-\omega^c) Q^c P^{d+1}\\ \
  [Q(t), Q(0)] &=&  (\omega^d-1)Q^{c+1} P^{d}
 \end{eqnarray}
 If we want Hermitian observables, we can take combinations like $P+P^{-1}$, which will have similar growth in commutators, that also use $M^{-1}$ instead of only $M$.
  
 When we take the absolute value, we find that these commutators grow with the typical size of the entries of the matrix $M$, 
 \begin{equation}
 |[P(t),P(0)]| \simeq |1-\omega^{|M|}|\simeq 2\pi |M|/N \label{eq:approx}
 \end{equation}
 which is valid  so long as $|M|<N$. If $M_1$ has eigenvalues $\lambda>1, 1/\lambda<1$, these entries grow like $\lambda^t$, so that the Lyapunov growth of commutators is controlled by 
 \begin{equation}
 \kappa_+=\log(\lambda),\label{eq:lyapcomp}
 \end{equation} 
 which is exactly the classical Lyapunov exponent. The Ehrenfest time will now be the time of saturation, which is controlled by $ \lambda^t/N<1$.
 Similarly, if we consider $Q^k P^{k'} (t)$ with $k,k'$ fixed of order $1$ in a condition where $N$ is large, the commutators with the basic fields  $(P,Q)$ or their inverses will grow typically like $2\pi ||k|\pm |k'||\lambda^t /N$, so the effective Ehrenfest time might be a little lower, but not by much.
 The behavior stops being classical when we go past the  time $\tau \simeq \log(N/2\pi)/\log(\lambda)$
 
 \subsubsection*{Averaging}
 
 Consider again the commutator $i[A(t),\CO(t_0)] $, where the factor if $i$ is to make sure that the operator is hermitian. If we average over a finite Hilbert space, we get that 
 \begin{equation}
 \tr( i[A(t),\CO(t_0)] ) =0
 \end{equation}
 so that on an average state it should vanish. What this means is that the commutator can not have a fixed sign and in general cancellations occur. Thus, the usual effect on a thermal system is that the response function 
 decays over time to equilibrium, when we average over a thermal state.

A way out of this quandary is obtained by squaring. In that case, we expect that for each semiclassical state the right hand side will grow exponentially, as twice the maximum Lyapunov exponent. Thus we would expect that 
\begin{equation}
\bra \psi ( i[A(t),\CO(t_0)])^2 \ket \psi \propto \exp(2\kappa_+(t-t_0))
\end{equation}  
 and the sign is positive. If semiclassical states are overcomplete (an usual assumption), then upon averaging over semiclassical states we should get that generically 
 \begin{equation}
\vev{\left ( i[A(t),\CO(t_0)])^2 \right)}\propto \exp(2\kappa_+(t-t_0))
\end{equation}  
because we are summing over quantities that all have the same sign and absolute value. There are two classes of time ordering in the expression. Time ordering where the $t$ and the $t'$ show together, and ordering where it is fully out of time ordering. The completely out of time correlation is given by
$\vev{A(t)\CO(t_0)A(t)\CO(t_0)}$. This is the out of time ordering correlator (OTOC) that  bounded the Lyapunov exponent by the temperature in the paper \cite{Maldacena:2015waa}.
  Again, the system saturates on a time scale of the order of the Erenfest time, when the OTOC decays, or the commutators saturate (when $A,\CO$ are bounded).
 
 Now, when we compute with $P,Q$, we should take the commutator $C$ and make it into a positive definite matrix. A natural choice is $C^\dagger C$. In this case, any extraneous phase cancels and we can compute directly that 
 \begin{equation}
 [P(t), P(0)]^\dagger  [P(t), P(0)] =| (1-\omega^b)|^2\simeq |2\pi b/N|^2\simeq \exp(2\lambda_+t)/N^2
 \end{equation}
 where we use the results from equation \eqref{eq:r1}, the same approximation as \eqref{eq:approx} and the same result that leads to the Lyapunov exponent definition \eqref{eq:lyapcomp}. 
 Again, the saturation time is the same as before.

 \section{Generalized cat maps}\label{sec:gqcm}

The cat maps were generalized to products of such vector spaces that can be put on a lattice, with a $P_\ell, Q_\ell$ at each site (with the same relations) in  \cite{Berenstein:2015yxu} and variables at different sites commuting with each other.

The idea is that if one has such a system then the following operations are an automorphisms of the algebra on the lattice
\begin{eqnarray}
P_i& \to &P_i  \nonumber\\
Q_i&\to& Q_i P_{i+1}\nonumber\\
Q_{i+1}& \to &Q_{i+1} P_i\nonumber \\
P_{i+1}&\to &P_{i+1}\label{eq:auto1}
\end{eqnarray}
The order in these don't matter because the different neighbor matrices commute. 
These are operations that entangle the nearest neighbors. This automorphism can be done for nearest neighbors on an arbitrary lattice, not just for points on a line. As such one can study  systems that exist in an arbitrary number of dimensions. For the purposes of this article we will restrict ourselves to one dimensional systems and for  certain more specific computations we will deal with a lattice with only two sites. 
 
 To build a generalized cat map as in \cite{Berenstein:2015yxu}, one can compose a local cat map at each site with a nearest neighbor entangler and after the system is constructed one iterates the automorphism. The system will also be determined by a $2 m\times 2 m$ matrix. Each $2\times 2$ block on the diagonal represents $P_i, Q_i$. The local cat map acts on each of these as a $2\times 2$ matrix, and the nearest neighbor entangler is a matrix that, at least for a lattice on a  line,  is near the diagonal giving rise to a banded matrix.
 The eigenvalues of this bigger matrix are the Lyapunov exponent of the system.  The reason for this is again that the action is a linear transformation on the logarithm of $P,Q$, and 
for such systems the matrix above represents the linear action on the coordinates. This means that in the classical limit $N\to \infty$ we recover the Lyapunov exponents easily.

 This is not the only way to proceed. To build more interesting systems, one considers more general automorphisms of the product algebra that send monomials to monomials. I will show a slight variation on the construction above where it is easier to compute analytically. One of the challenges for computations above is that in the cat map automorphism one has phases, and one has to further iterate. So for the purposes of computing, it is easier if one can find a setup where there are no phases that need to be addressed.
 The idea for the variation on the theme above is that the label $P,Q$ are somewhat arbitrary. We can easily change the names by taking $P\to Q, Q\to P^{-1}$. Consider now an automorphism defined by
 \begin{eqnarray}
 Q_1&\to& Q_1^a Q_2^b \nonumber \\
 Q_2&\to& Q_1^c Q_2^d \nonumber\\
 P_1&\to& P_1^{\tilde a} P_2^{\tilde b}\nonumber \\
  P_2&\to& P_1^{\tilde c} P_2^{\tilde d} 
 \end{eqnarray}
  where all the powers are integer. Clearly the images of $Q_1, Q_2$ commute with each other, and so do the images of $P_1,P_2$ and there is no ordering ambiguity. One can easily check that the condition for this map to  be an 
  automorphism is that 
  \begin{equation}
  \begin{pmatrix}
  a&b\\
  c&d
 \end{pmatrix}  \begin{pmatrix}
  \tilde a & \tilde c\\
  \tilde b &\tilde d
 \end{pmatrix}=  \begin{pmatrix}1&0\\
  0&1
 \end{pmatrix}
 \end{equation}
 Since all of the elements of the matrix are integers, we must have that $\det(M)=\pm1$. We will restrict to those matrices such that $\det(M)=1$ for simplicity.

 Notice that the map with 
 \begin{equation}
M= \begin{pmatrix}
 a&b\\
 c&d
 \end{pmatrix}=\begin{pmatrix}1&1\\
  0&1
  \end{pmatrix}
\end{equation}
corresponds exactly to the nearest neighbor entangler discussed initially, but with a relabeling of the $P,Q$. The automorphisms detailed here contain as a special case the automorphism \eqref{eq:auto1}. However, they are more general and can produce non-trivial eigenvalues that lead to chaos already on their own. Thus, we do not need the additional cat automorphism on each lattice site to induce chaos in the system.

Iterating over a general $M$ produces a cat map dynamics on the $Q$, and the `inverse' cat dynamics on the $P$. The dynamics on $P$ is actually built from the inverse transpose, but that has the same eigenvalues as the inverse of $M$.
 Such a system can be generalized to many $P,Q$ matrices, schematically of the form
\begin{eqnarray}
Q&\to& Q^M\nonumber\\
P&\to& P^{(M^{-1})^{ T}}
\end{eqnarray}
where $M \in SL(\ell,\BZ)$, where $\ell$ is the total number of sites. The matrix will be nearest neighbor-like if both $M$ and $M^{-1}$ is sparse. As noted in \cite{Berenstein:2015yxu},  models with nearest neighbor properties also mimic the Lieb-Robinson bound \cite{lieb1972} for propagation of information and thus can in principle serve as toy models for relativistic field theories (they have the equivalent of a speed of light).
 
Now, in this setup,  the eigenvalues of $M$, together with the eigenvalues of $M^{-1}$ are the Lyapunov exponents of the full system. The advantage is that there are no normal ordering ambiguities in the maps, so everything can be studied in the absence of phases.
This is the setup that will be explored in the current paper. Obviously additional generalizations where on each $P,Q$ one performs an additional cat map are also interesting, but they require determining  the phases in a convenient way for computations and iterations.
The main question we are interested in is to study how the different quantum tori entangle with each other when we start from a product state for a variety of different classes of initial states. 
  
Our interest is to study initial states that are factorized on the sites. In principle these are easy to prepare. As we evolve the system, the expectation values of the variables $P,Q$ at a later time can be determined from the map on the algebra, rather by evolving the state 
directly. Because the map sends monomials to monomials, at a later time these expectation values are factorized in the initial state. As such, their computation is easy, even for large lattices. This suggests that even in the absence of analytical results, these systems are very amenable to  numerical implementations to study the dynamics for general (factorized) initial states.

\section{Two site problem and semiclassical entanglement}\label{sec:twosite}

Let us start with the simplest setup. What we can call a two site problem. That is, a collection of two sets $Q_1,P_1$, $Q_2,P_2$ that are entangled with a cat map. For simplicity let us now consider a Unitary transformation associated to the matrix
\begin{equation}
M=\begin{pmatrix}2&1\\
  1&1
  \end{pmatrix}
\end{equation}
the Arnold cat map. We will denote this by a graph 
\begin{equation}
\xymatrix{1 & 2\\
1\ar[ur]&  2\ar[ul]
}
\end{equation}
where the crossing indicates that the two are getting mixed (by $M$).
The dynamics results from iteration
\begin{equation}
\xymatrix{1 & 2\\
1\ar[ur]&  2\ar[ul]\\
1\ar[ur]&  2\ar[ul]\\
1\ar[ur]&  2\ar[ul]
}
\end{equation}
What we want to do now is understand a little bit better the map. Notice that in the map
\begin{equation}
Q_1\to Q_1^a Q_2^b
\end{equation}
the right hand side term commutes with the one on the left. This is also true when we consider the image of $Q_2$ (they also commute with $Q_2$). That means that both $Q_1(0),Q_2(0),Q_1(t), Q_2(t)$ can be diagonalized simultaneously. 
We can imagine the map $U_M$ acting on a pair of kets $\ket{k}\otimes \ket {k'}\simeq \ket{k,k'}$ which are eigenvalues of $Q_1, Q_2$. After the action, we are supposed to evaluate the eigenvalues of the images of $\tilde Q_1, \tilde Q_2$ acting on the state, and think of them as $Q_1, Q_2$ acting on the evolved state. This is equivalent to sending
\begin{equation}
\ket{k,k'} \to \ket{ a k + bk', ck+d k'} \label{eq:actionstates}
\end{equation}
rather than doing the map on the algebra. 
The upshot is that we end up acting like matrix multiplication on the labels of the state. The multiplication is modulo $N$, so it is a discretized version of the cat map on a regular torus (the action is the action on the  rational points of the torus $T^2$ with denominator $N$). 

That is, the dynamics can be thought of as being given by a permutation dynamics on a particular basis of  states. This is particular to all the maps that send $Q\to Q^M$ that we are discussing in this paper. This insight is what will let us do a lot of computations, not only on this two site problem, but also on a multi-site lattice.

For the two site problem, our first objective is to study how the two factors of the Hilbert space get entangled with each other as time goes by, when we start with a (pure) product state and to study the dependence on the choice of states. The initial states this way can be described by a  wave function 
\begin{equation}
\psi\simeq \sum_{k,k'}\psi(k,k')\ket{k,k'}=\sum_{k,k'} \psi_1(k)\psi_2(k') \ket{k,k'}
\end{equation}

We will do various such analysis. 

\subsubsection*{The basis states}

The simplest states that we can consider are basis states in the basis we have chosen. These are product states of the individual basis and are pure on each factor. Remember that 
\begin{equation}
\ket{k,k'}=\ket{k}\otimes \ket {k'}
\end{equation}
 As we evolve the state, the state gets permuted into other another state that has the same properties. There is no entanglement produced between the different Hilbert spaces. These are $N^2$ states that produce no entanglement, even though they evolve in time.
 In these setups the entanglement entropy satisfies $s_1=0=s_2$ for all times. A full basis of states experiences no entanglement dynamics.

\subsubsection*{A basis state times a random state} 
 
Consider now the states of the form $\psi =\ket k \otimes  \sum_{k'} \psi_2(k') \ket {k'}$. These will evolve into a wave function of the form
\begin{equation}
\sum_{k'}  \psi_2(k') \ket{M^t\cdot(k,k')}\label{eq:evoldisc}
\end{equation}
where we need the $t$ powers of the matrix $M$.
 Let us now trace over the second Hilbert space. We care to know if $M^t\cdot(k,k')$ is different in both the first and the second coordinate between the different choices for $k'$. This depends on the prime factorization of $N$. We will assume in what follows that $N$ is prime as an illustration.  Our problem in the second coordinate reduces to the problem of understanding if the $t$  power of $M$ is such that the matrix component $ (M^t)_{22}\neq 0 \mod(p)$ is a multiple of $p$ or not.
 
For $p$ large and small values of $t$, this is so always, so  it has an inverse modulo  $p$ and all the second coordinates are distinct. This also applies to the first coordinate if at the same time $(M^t)_{21}\neq 0 \mod(p)$.  What this means is that the different $k'$ states end up being imaged into different positions both in the first and second coordinates. 

That is, the wave function looks like
\begin{equation}
\sum_{\tilde k = f( \tilde k')} \tilde\psi(\tilde k, \tilde k')\ket{k,k'}
\end{equation}
 with no repetitions of any $\tilde k,\tilde k'$, but only one term with each possible value of $k$ and one term with each possible value of $k'$. When we trace over the second coordinate, we thus get a diagonal density matrix in the basis representation, whose entries are given by
 \begin{equation}
 |\psi(k')|^2
 \end{equation}
 for some $k'$. That is, after one shot, the density matrix is diagonal in the $k$ basis, and the coefficients are the probabilities for being in the different $k'$ in the initial state. We thus erase the phase information for $\psi(k')$ (we can not recover it from observations of $Q_1,P_1$). The entanglement entropy is then 
 \begin{equation}
 s= -\sum |\psi(k')|^2\log(|\psi(k')|^2) \label{eq:entropyt}
 \end{equation}
 and it is constant until the point when the condition of independence of the two coordinates fails. When it fails, the system is factorized again and it has a large dip in the entropy to zero. 
For a random state, the entropy is of order $\log(N)$ in one shot and is essentially determined by the size of the matrices.
For other states, where we choose the $\psi_2(k)$ to be non generic the numbers will change. For example, if we require low uncertainty on $k'$ we will have a low entropy.
 
\subsubsection*{The dual basis states}

Just like the images of $Q$ commute with the $Q$ in the initial time, the images of $P$ commute with the $P$ in the initial time. The same arguments that lead to no entropy being generated by the action of the cat map are true when the initial state is a product state in the $P$ basis rather than the $Q$ basis. That is, we have a second basis where there is no entanglement production: the Fourier transform basis.
As such, the properties of the system can be analyzed in either of these two perspectives with similar results. This statement about the entropy would be surprising in the $Q$ basis, as each such state has probability $1/N$ for being in each of the different $Q$ eigenstates (for each $Q$).
A generic such state should go to entropy $\log(N)$ in one shot, similar to the basis state times a random state. 
Indeed, if the initial state is a $Q$ eigenstate in the first entry and a $P$ eigenstate in the second entry, then the entropy is $\log(N)$ after one shot. To cancel the entropy production, we need specific linear combinations of $Q$ eigenstates. This suggests that one can reduce entropy production by using interference phenomena in quantum mechanics. 

\subsubsection*{Gaussian states and entropy production controlled by Lyapunov exponents.}

A third class of states that is interesting is to consider states that minimize the combined uncertainty for $P,Q$ simultaneously on each site. To define such states, we first need a useful definition of the uncertainty of $P$ that is a unitary, rather than a Hermitian operator.
Consider the following expression
\begin{equation}
\vev{(P^\dagger - \exp(-i\theta))(P-\exp(i\theta))}_\psi \geq 0
\end{equation}
 which is an expectation value of a positive operator. This vanishes for values of the phases and states where $P \ket\psi = \exp(i\theta) \ket\psi$ is an eigenstate of $P$. Such states have no uncertainty in $P$. We define the uncertainty for $P$ as follows
 \begin{equation}
\Delta P^2 = \min_\theta \vev{(P^\dagger - \exp(-i\theta))(P-\exp(i\theta))}_\psi 
 \end{equation}
We say a packet is centered at the origin if the angle that minimizes the value is at $\theta=0$. Similarly, we define
\begin{equation}
\Delta Q^2 = \min_\eta \vev{(Q^\dagger - \exp(-i\eta))(Q-\exp(i\eta))}_\psi 
\end{equation}
and say that a wave packet is centered at $0$ if the $\eta$ that  minimizes the expression is $\eta=0$. We say we have a minimal uncertainty packet located at $(\theta= 0,\eta=0)$ if the wave function $\ket\psi$ is such that for fixed $A,B >0$, the wave function minimizes the combination
\begin{equation}
A \Delta P^2_\psi + B \Delta Q^2_\psi =\vev{ A(2-P-P^\dagger)+ B(2-Q-Q^\dagger)}_\psi  \geq 0
\end{equation}
Notice that the right hand side of the expression is an expectation value of a Hermitian operator $\hat H=  A(2-P-P^\dagger)+ B(2-Q-Q^\dagger)$. Thus, if we consider all possible $\psi$, the right hand side is minimized if $\ket \psi$ is exactly the eigenvector of the Hermitian operator with the smallest eigenvalue. For simplicity we choose $A=B$ to make a symmetric state between $P,Q$, but we can adjust this. 
 This is very similar to the procedure for finding points in a fuzzy geometry described in \cite{Ishiki:2015saa}, but without taking the dimension of the matrices to be large. 
 The  problem for $\ket \psi$ can be solved numerically, by looking for the eigenvector of $P+P^\dagger+ Q+Q^\dagger $ with maximum eigenvalue. Numerically,
 the maximum and the minimum eigenvalue of this combination have very close norm $(P, Q \simeq \pm 1)$, so that it is easier to use 
$\hat \Gamma=P+P^\dagger+ Q+Q^\dagger +\alpha$ for $\alpha$ a small positive number. 
This way we can iterate $\psi \to \hat \Gamma \psi$ and normalize $\psi$ until the state converges. This lets us produce easily wave-functions with the required characteristics.

Notice that in our presentation $P\simeq \exp(i x_1), Q\simeq \exp(i x_2)$, we get that for a packet centered at zero
\begin{equation}
P = 1+ i x_1 -\frac {x_1^2}2 +\dots
\end{equation}
so that 
\begin{equation}
\vev{1-P} \simeq \frac {\vev {{x_1^2}}}2 \simeq \frac 12 \Delta x_1^2
\end{equation}
And similarly $\vev{2-P-P^\dagger} \simeq  \Delta x_1^2$ so that what we have labeled the uncertainty on $P$ is equivalent to a good approximation to the uncertainty of $x_1$. Similarly for $x_2$. If the uncertainty of $x_1, x_2$ are both small, we expect that we can't tell if $x_1$ and $x_2$ are periodic or not. As such, minimizing the uncertainty should give us a state that is approximately very similar to the ground state of a harmonic oscillator (remember that $x_1, x_2$ are conjugate variables). Such a state will be essentially a Gaussian distribution centered around the location in phase space where we have centered the wave packet. The evolution of such a state should follow the entropy production of a regular Gaussian state \cite{Asplund:2015osa} to a good approximation, until the point where the uncertainty is large enough to start finding that $x_{1,2}$ are periodic variables \cite{Berenstein:2015yxu}. That is, we should expect that the entropy production is semiclassical in this regime and that it grows with a rate given by the sum of the two largest Lyapunov exponents of the system (see also \cite{Bianchi:2017kgb}). 

We want to compare the evolution of the entropy of the system with a different system,  where we ignore the periodicity of the $x_i$ variables. In this modified system the dynamics is given by iterated linear transformations of the $x_{1,2}^i$ into each other. That evolution is a model for what the entropy production should look like up to the point where the spread of the wave function is comparable to the periodicity of the torus.
The entropy of a density matrix for a Gaussian state (for a set of $k$ oscillator degrees of freedom) can be computed as 
\begin{equation}
\sum_{s=1}^k g(\nu_s)
\end{equation}
where the function $g$ is given by
\begin{equation}
g(x)= \frac{x+1}2 \log\left(\frac{x+1}2\right) -\frac{x-1}2 \log\left(\frac{x-1}2\right) 
\end{equation}
and the $\nu_k$ are the symplectic eigenvalues of the covariance matrix 
\begin{equation}
\Delta^2_{ij} = \vev {\frac 12(x^i x^j+x^j x^i)} 
\end{equation}
in the Williamson form \cite{GaussQI} (the $x^i$ here have normalized canonical commutation relations) and the $x^i$ are the phase space variables that have not been traced over: we trace over the second variable $x_1^2, x_2^2$.
These eigenvalues are the eigenvalues of $\Delta^2_{ij} \Omega^{jk}$ where $\Omega$ is the symplectic form.

For a single degree of freedom $\nu^2$ is the determinant of the covariance matrix. This is the product of the uncertainties.  To normalize the variables, we know that he initial state has entropy equal to zero (a product state), which corresponds to $\nu=1$. 
The evolution of the  initial Gaussian state (with the same uncertainties on all variables) will produce a diagonal  covariance matrix
\begin{eqnarray}
\Delta^2  x_1^1(t) = (a^2_t +b^2_t)\Delta^2  (x_1^1)(0) \label{eq:deltax}\\
\Delta^2  x_2^1(t) = (c^2_t +d^2_t)\Delta^2  (x_2^1)(0)\label{eq:deltap}
\end{eqnarray} 
so that at time $t$ we compute that 
\begin{equation}
\nu_t^2 = (a^2_t +b^2_t)(c^2_t +d^2_t) \nu(0)^2=(a^2_t +b^2_t)(c^2_t +d^2_t)
\end{equation}
where we have used $\nu(0)=1$.
This is our theoretical expectation for the value of the entropy at time $t$, which can be readily evaluated from the powers of the matrix $M$.
 We can also compute directly the entropy from the (numerical) wave functions that minimize the uncertainty. 
 
\begin{figure}[ht]\begin{center}
\includegraphics[width= 6 cm]{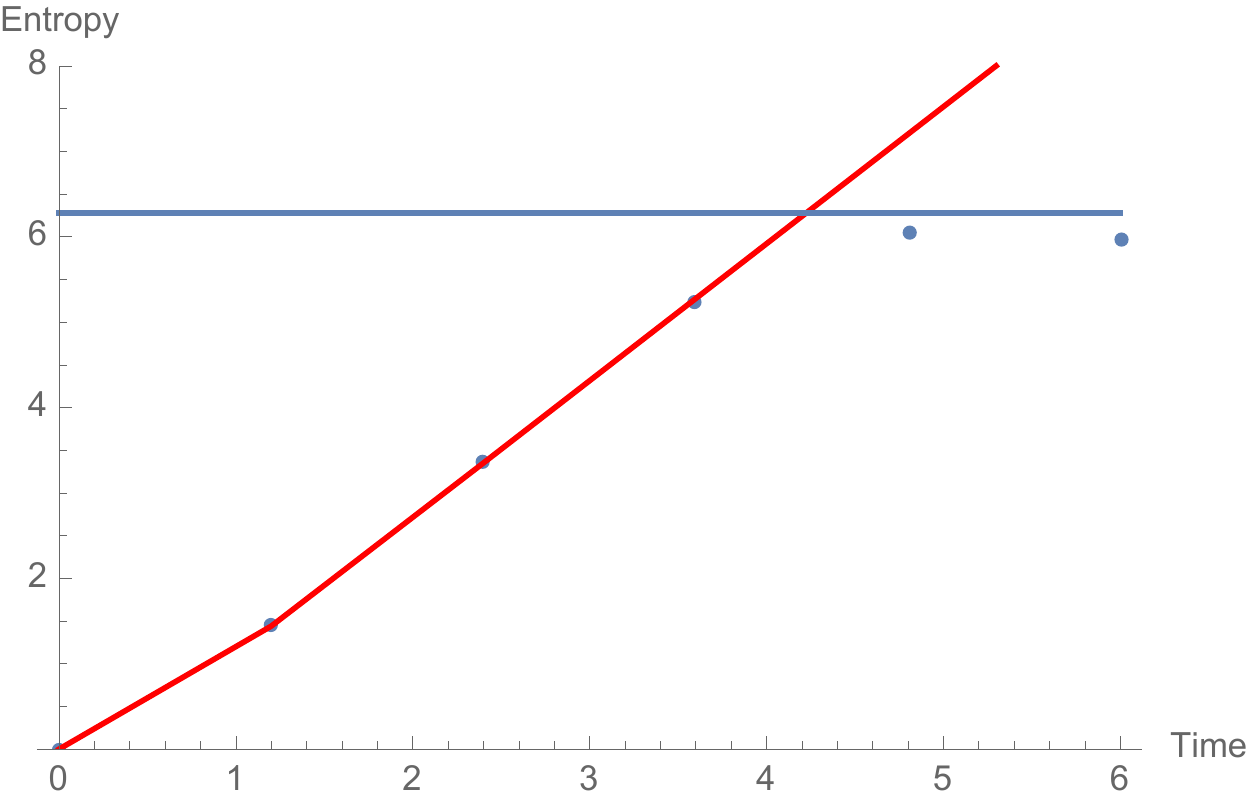}\caption{Evolution of entropy over time for a minimum uncertainty packet. The solid line is the theoretical Gaussian estimate, the dots are the results of the numerics for the density matrix, and the horizontal line represents the maximum allowed value of the entropy for a site.  }\label{fig:entropevol}
 \end{center}
\end{figure}

 An example is plotted in figure \ref{fig:entropevol}. As can be seen from the graph, the Gaussian estimate is very accurate until close to saturation of the entropy. 
 
The asymptotic slope of the Gaussian estimate is exactly the Kolmogorov-Sinai entropy of this system \cite{Asplund:2015osa, Bianchi:2017kgb}, which is twice the  logarithm of the largest eigenvalue of the matrix $M$.

We conclude that the details of how the system gets entangled over time depends very much on the state that is being discussed. We expect that for sufficiently semiclassical initial states, the entanglement is going to proceed as the Gaussian estimate until saturation. On the other hand, very quantum states (random states, or states that are eigenstates of $P,Q$) evolve very differently. One can have either no entropy production (this is probably due to the fact that the $P,Q$ are not mixing and is very special to this system) or saturation of the entropy in one shot, which is probably the generic result for most systems. What is important is that  all these cases we have analyzed have a good theoretical control of the entropy production (or lack thereof). For other states that are in between classical and quantum, one  probably needs to resort to simulations to get a better understanding of the physics.

\section{A family of one dimensional lattices as a one way communication channel}\label{sec:CNOTqft}

Our next example is to set up a one dimensional lattice, with nearest neighbor entanglers  whose dynamics can be encoded by  the matrix
\begin{equation}
M=\begin{pmatrix} 1&1\\
0&1\end{pmatrix}
\end{equation}
in the following pattern
\begin{equation}
\xymatrix{&&&&&&\\
\dots &&\ar[ur] & \ar[ul]&\ar[ur] & \ar[ul] &\\
&6 \ar [ur] & 5 \ar[ul] &4 \ar [ur] & 3\ar[ul] &2\ar[ur] &1\ar[ul] 
}
\end{equation}
The full matrix $M_{total}$ will be of the form
\begin{equation}
M_{total}= \begin{pmatrix} \ddots&&&&&\\
&1&1&& &\\
&0&1&&&\\
&&&1&1&\\
&&& 0&1&\\
&&&& &1  \end{pmatrix}
\begin{pmatrix}  \ddots&&&&&\\
&\ddots &&&&\\
&&1&1&&\\
&&0&1&&\\
&&&&1&1\\
&&&& 0&1
\end{pmatrix}=\begin{pmatrix}  \ddots&&&&&\\
&1 &1&1&0&0\\
&&1&1&0&0\\
&&0&1& 1&1\\
&&&&1&1\\
&&&& 0&1
\end{pmatrix}
\end{equation}
We will follow the convention that the sites $1, 2, \dots$ are represented in reverse order, so that the matrix $M_{total}$ is upper triangular. 
This dynamics is done with the entangler that is equivalent to the one found in \cite{Berenstein:2015yxu} as an automorphism of the algebra in equation \eqref{eq:auto1}, but with labelings so that the $Q$ and $P$ don't mix.

Because the matrix is upper triangular, one immediately finds that all the eigenvalues are one, which suggests that the system is not chaotic.
As given, the system has a well defined left and right, which we call boundary conditions. 

We can also choose periodic boundary conditions. For example, consider four sites, and then 
\begin{equation}
\bar M_{4\times 4}= \begin{pmatrix} 1&&&\\
&1&1&\\
& 0&1&\\
1 && &1  \end{pmatrix}
\begin{pmatrix} 
1&1&&\\
0&1&&\\
&&1&1\\
&& 0&1
\end{pmatrix}
\end{equation}
where the bar indicates that we have chosen periodic identifications. It is easy to check that the eigenvalues of $\bar M_{4\times 4}$ are not unitary numbers so the periodic system is chaotic.

We find this way that one of the simplest systems for  the lattice  dynamics is chaotic or not depending on the periodicity conditions that we impose and that the system with periodic boundary conditions is chaotic. We will study the non-periodic system, which should already have some traces of chaos in it. After all, in the middle of the chain we would not know if the system is periodic or not for a long time.

\subsubsection*{The Q-basis dynamics}

Just as in the example studied in the previous section, we find that the various $Q$ commute with each other at later times, and also with the initial $Q$ variables, so that they can be diagonalized simultaneously. A basis will thus be described by a collection of integers 
$\ket{\dots k_2, k_1}$ with $k_i$ integers modulo $N$. We will collectively call these $\ket{\vec k}$. Our dynamics will be similar to the one expressed in equation \eqref{eq:evoldisc}, and will go as follows
\begin{equation}
U_t \ket {\vec k} = \ket{M^t_{total}\cdot \vec k}
\end{equation}
where the entries of the images  $\vec k$ are all evaluated modulo $N$ and we end up with the powers of the matrix $M$. This is exactly like the linear dynamics on a rations sublattice of a torus (with denominators equal to $N$).

A basis state goes to a basis state. Thus, there is no entanglement entropy production for the evolution of a single basis state. 

Notice also that the state $\ket {\vec 0}$ goes to itself under the evolution and can be considered as a {\em vacuum} for the dynamics.

Consider a state that is given by vacuum everywhere except near the tail, where the state at site $1$ is set to $s$ rather than zero. Notice that under the evolution this number will not change, because $Q_1\to Q_1$. What we see is that under the evolution we have that 
\begin{equation}
\ket{\vec 0, s} \to \ket{\vec 0,  s, s,s} \to \ket{\vec 0,  s, s ,  3s,  2s, s} \to   \ket{\vec 0,  s, s ,  5s,  4s, 6 s, 3s,  s}\to \dots 
\end{equation}
so the letter $s$ is being sent to the left at speed $2$, while the last letter stays constant. We can think about this as sending a message to the left. 

Now, we need to analyze the dynamics of the system modulo $N$, and here it matters what precise value of $N$ we choose. We will choose the smallest non-trivial value first $N=2$. In this case the 
primitive entangling operation between two nearest neighbors is a two q-bit gate. One can easily recognize it as the a controlled not gate (CNOT), that acts as follows
\begin{eqnarray}
CNOT \ket{00} &\to &\ket{00}\\
CNOT \ket{01} &\to & \ket{01}\\
CNOT \ket{10} &\to & \ket{11}\\
CNOT \ket{11} &\to & \ket{10}
\end{eqnarray} 

As written the dynamical system we have described can be (easily) programmed into a quantum computer, so long as CNOT gates are implemented efficiently. This is a special example of a controlled phase gate (current architectures suggest a  fidelity per gate operation of $0.98-0.99$ \cite{Martinis}).  Because the evolution of the basis states can be thought of in terms of a classical update on a  discrete lattice with local updates, one can think of the system as a quantum version of a cellular automaton.

\begin{figure}[ht]\begin{center}
\includegraphics[width=6 cm]{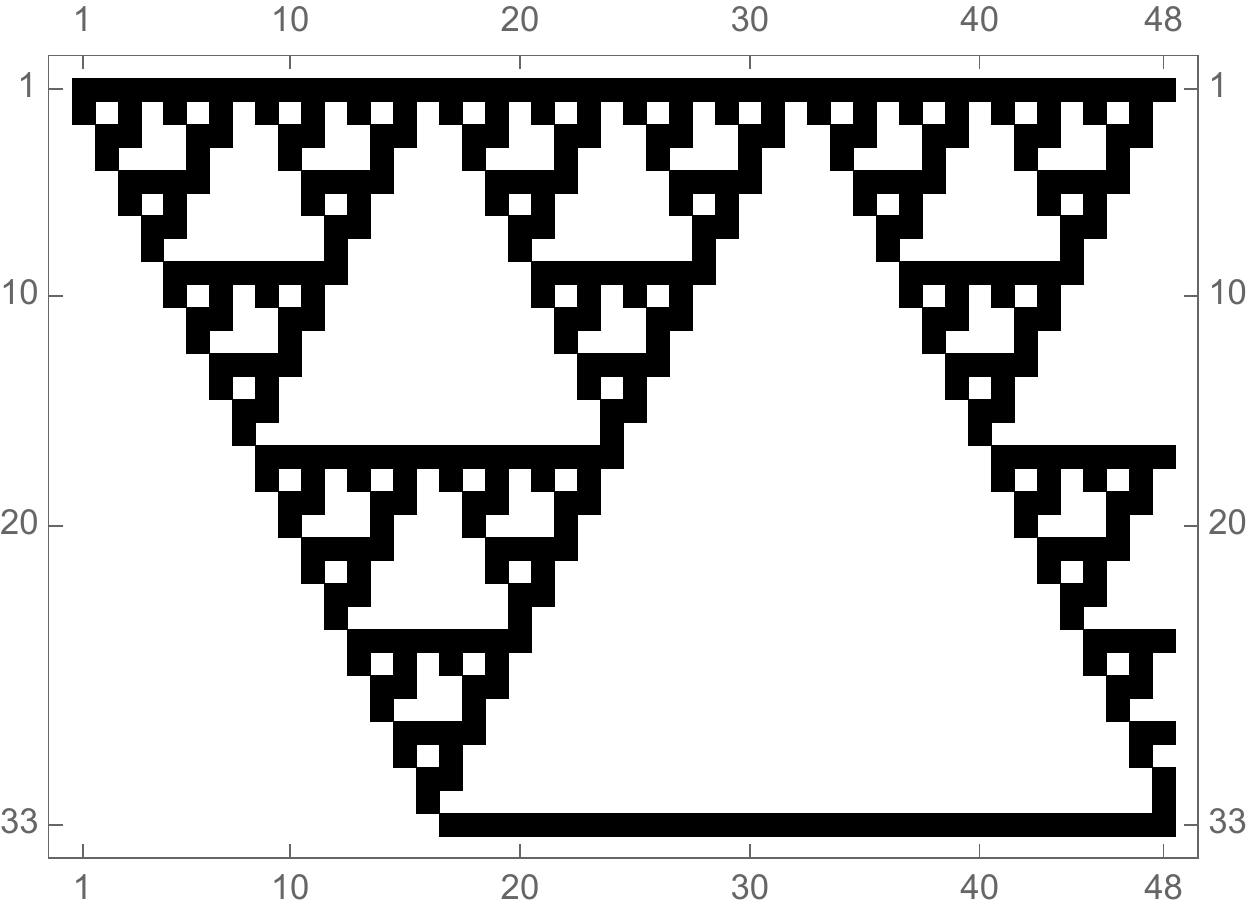}\caption{ Encoding of the first bit at various positions in time. The position is shown vertically (going downward) and discrete time runs to the right }\label{fig:Sierp}
 \end{center}
\end{figure}

Our goal now is to understand the dynamics in this special case. We find that the bit at the end is encoded in the bit at position $w$ at time $t$ if the component $M^t_{w1} =1$. This is shown in the figure \eqref{fig:Sierp}.

The figure seems related to Sierpinsky's triangle, a well known fractal. The system is periodic in the horizontal direction for each position, but at each position further down the periodicity is longer.  
The position one is the one that keeps the message fixed in time (the coordinate $Q_1$ is fixed). The idea is that the message in the bit one is sent periodically in time. 
In a coarse grained sense, the number of hits per unit time at position $s$ decays with $s$, scaling with $s$. This suggests some  connection with critical behavior. 

\begin{figure}[ht]\begin{center}
\includegraphics[width= 6 cm]{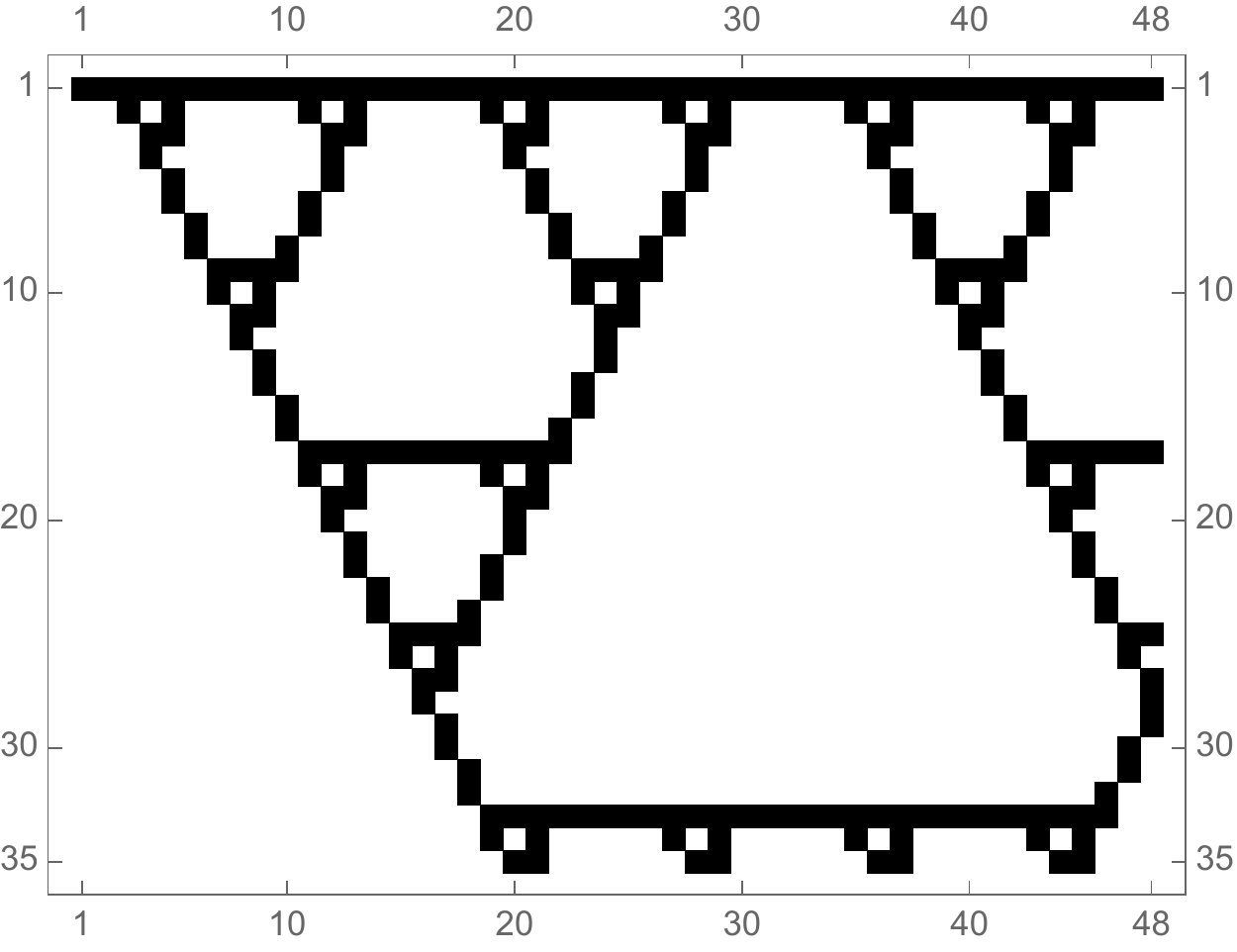}\caption{ Encoding of the first 5 bits at various positions in time. The position is shown vertically (going downward) and discrete time runs to the right }\label{fig:Sierp5}
 \end{center}
\end{figure}

\begin{figure}[ht]\begin{center}
\includegraphics[width= 6 cm]{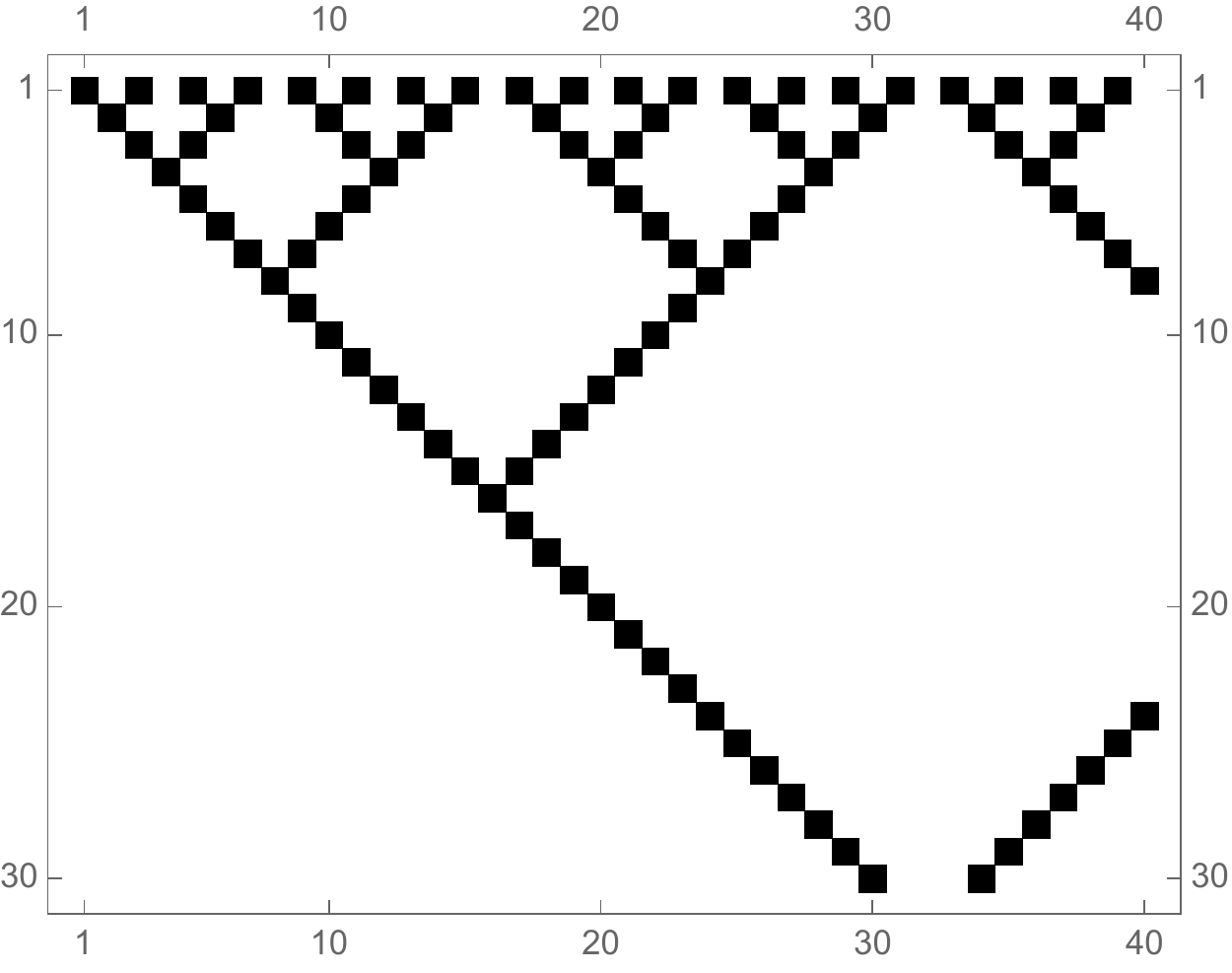}\caption{Time of exit of a message of length $k$, starting at the $k+1$ site. Time runs horizontal and faithful encoding of message of length $k$ starting at the $k+1$ site is shown vertically.  }\label{fig:exittime}
 \end{center}
\end{figure}

Similarly, we can consider the problem of sending $k$ bits of information in a message, and reading the message at position $s$, by reading the next $k$ bits  that begin at that position (for a message of length $2$ we start at position $1$, but consider the bits $1$ and $2$, for a message of length $3$ we start at position $1$, including $1,2,3$, etc). To check if the message is encoded faithfully, we need to look at the part of the matrix $M^t_{total}$ that maps the initial $k$ initial bits to the target $k$ bits starting at $s$.  Schematically, we need to look at a submatrix
\begin{equation}
M^t_{tot}\simeq \begin{pmatrix}
&&&\\
&&&\begin{pmatrix}\cdot & \cdot\\
\cdot & \cdot\end{pmatrix}\\
&&&\\
\end{pmatrix}\begin{pmatrix}\\
\\
\cdot\\
\cdot
\end{pmatrix}
\end{equation}
that maps the initial bits to the desired bits. In the simplest setup this is a square matrix. The message is faithfully encoded (in that every initial message is sent to a different message at the desired position in time)  if the kernel of the submatrix vanishes. This can be tested by evaluating the determinant of the relevant submatrix (modulo $N$) as a function of time.  An example with a message of length $5$ is shown in figure \ref{fig:Sierp5}.
The pattern is very similar to the one found previously, with a strong resemblance to Sierpinsky triangle, but it is also thinner (less dense). There is a delay for the message to get out, but once it does it travels at speed $2$ again. 
It is easy to check numerically that a message of length $k$ gets out in time at most $k$, as shown in figure \ref{fig:exittime}.
As the figure shows, the time to send a message of length $k$, faithfully to the next $k$ bits is exactly $k$. The structure of the message copying is also self-similar, with the branching points at the powers of two.
Similar results are found when $N$ is varied. 
The fact that the matrix $M_{total}$ is upper triangular means if the first $s$ sites  are initialized at zero, they stay fixed at zero for all time, and only the message of the next $k-s$ bits is sent periodically. Thus, we can forget these empty sites to the right if we want to. 

Now that we understand that the (classical) message of length $k$ is copied in time $k$, we can ask what  this means for entanglement entropy production. Clearly, once an initial state is classically copied with an encoding as follows
\begin{equation}
\ket{\vec 0, \vec \alpha} \to \ket{\vec 0, ( \vec \alpha), (\vec \alpha) } 
\end{equation}
where the parenthesis indicates some faithful encoding (without being too specific), the encoded message keeps traveling to the left and can be identified as two correlated copies of the original (classical) message. When we ask what happens to the rightmost $k$ cells, we find that  the density matrix of those cells needs to be diagonal in the $Q$ position basis. This is because the different encodings $(\vec \alpha)$ of the traveling message are orthogonal. In a sense, we find that the evolution is (unitarily) equivalent to 
\begin{equation}
\ket{\vec \alpha} \to \ket{\vec \alpha}\otimes  \ket{\vec \alpha}
\end{equation} 
The maximal entropy generated this way is $S_{max}= k \log(N)$, and this occurs in time $k$ (at most), so the maximal entropy production per unit time is bounded below by $\log(N)$. This is the entropy of a single site and it is independent of the possible Lyapunov behavior of the system. The density matrix for the last $k$ sites will have as entries the probabilities for being in each of the classical $Q$ states in the initial wave function (which get permuted over time in the dynamics), just like in equation \eqref{eq:entropyt}. This density matrix will also forget the phases of the wave function.
When we evolve the system further, multiple copies of the message get generated, so the evolution will produce states that evolve schematically  (up to precise details of the encodings) as 
\begin{equation}
\ket{\vec \alpha} \to \ket{\vec \alpha}\otimes  \ket{\vec \alpha}\to \ket{\vec \alpha}\otimes  \ket{\vec \alpha}\otimes  \ket{\vec \alpha}
\end{equation} 
These will produce multi-partite entanglement in a generalized GHZ state \cite{GHZ} when we take linear combinations of the initial messages to build the quantum message. 

Notice that the rate $\log(N)$ is similar to the maximal rates that are available in systems with a  Hamiltonian evolution \cite{bravyi2007upper,Acoleyen}. Since the speed of the message is twice as fast, we conjecture that the maximal rate of entanglement that is possible in this system is actually 
\begin{equation}
\frac{\Delta S}{\Delta t} \leq 2\log(N) \label{eq:qbound}
\end{equation}
 per unit time.

\subsection*{The P vacuum stops all messages}

Consider now the problem of studying the system in the $P$ basis, rather than the $Q$ basis. 
The problem will now use the matrix
\begin{equation}
(M^{-1})^T=\begin{pmatrix} 1&0\\
-1&1\end{pmatrix}
\end{equation}
as the main building block. 
\begin{equation}
(M^{-1})^T_{total}= \begin{pmatrix} \ddots&&&&&\\
&1&0&& &\\
&-1&1&&&\\
&&&1&0&\\
&&& -1&1&\\
&&&& &1  \end{pmatrix}
\begin{pmatrix}  \ddots&&&&&\\
&\ddots &&&&\\
&&1&0&&\\
&&-1&1&&\\
&&&&1&0\\
&&&& -1&1
\end{pmatrix}
\end{equation}
which is now {\em lower triangular}. So, in analogy with our previous analysis in the $Q$-basis, this is a system that sends classical information messages moving only to the right. 
This means that if the left $M$ sites are in a P-basis state (in a $P$ eigenstate), the sites will stay in a $P$ eigenstate as they evolve, and no entropy is generated. This is regardless of what we do on the rightmost sites. 
What this means is that if the left sites are in a $P$-eigenstate, we can stop the messages that are trying to be emitted from the right: there is no entanglement between the left and the right. 
In the particular case of $N=2$, such $P$ eigenstates are Hadamard product states and should be easy to engineer on a quantum device.

Similarly, we can set up the system so that all the entangling dynamics happens in a small interval: the rightmost states are in a $Q$ eigenstate, the leftmost states are in a $P$ eigenstate, and the ones in the middle evolve in a way that 
entangles them, but where the entanglement stays localized in the middle region.
It is important in this analysis that the states be in the $P$ eigenstate all the way to the left. Otherwise, if it is only in a region and then it reverts to the $Q$ ground state, the analysis in the previous section shows that the classical messages get out if we wait long enough. This means that the rightmost states will get entangled with other sites eventually. 

For $N=2$, the states that stop the entangling between regions are Hadamard statest
\begin{equation}
\ket \psi_L = \frac{\ket 0 +\ket 1}{\sqrt 2}
\end{equation}

\subsection*{Traces of chaos}

The last thing we can do is to understand how Gaussian states grow (locally at one site), before the system has time to find out that it is in a finite box. There are two locations where it is interesting to analyze: the dynamics. At one end, and in the middle of the chain. We will do 
the calculation assuming that we can ignore the finiteness of the local site to estimate the entropy growth. Essentially, we will posit that the entropy growth is given as in the solid lines in figure \eqref{fig:entropevol}, which follows from ignoring the finiteness of $N$.

The equations are going to be similar to equations \eqref{eq:deltax},\eqref{eq:deltap}. They follow from 
\begin{equation}
x^k_1(t)= (M_{total} ^t)^k_j x^j_1(0)
\end{equation}
so that 
\begin{equation}
\Delta^2 x^k_1(t) = \left[(M_{total} ^t)(M_{total} ^t)^T\right]_{kk}\Delta^2 x^k_1(0)
\end{equation}
For the end site, we find that since $M$ is upper triangular, that $\Delta^2 x^1_1(t)=\Delta^2 x^1_1(0)$. This is the statement that $Q_1(0)$ is conserved. 

\begin{figure}[ht]\begin{center}
\includegraphics[width= 6 cm]{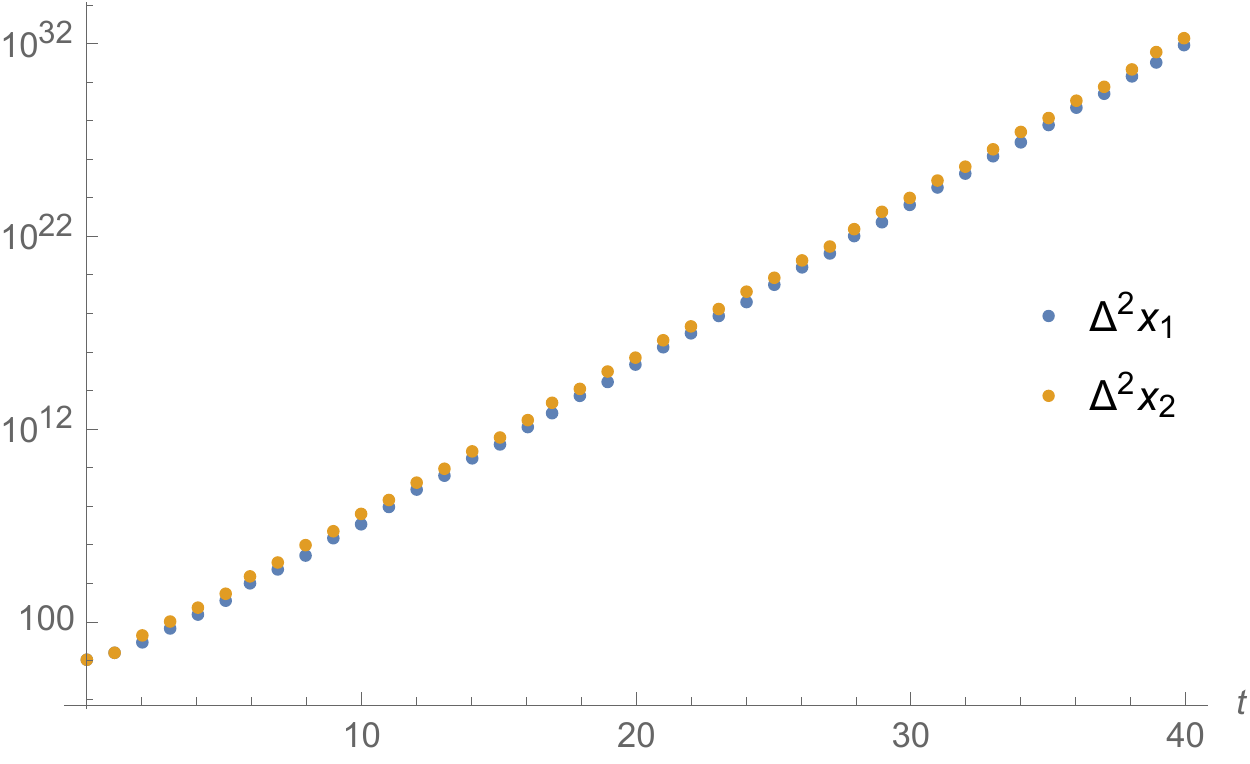}\caption{Evolution of the square of the uncertainty $\Delta^2 x^k(t)/ \Delta^2 x^k(0)$ over time for a minimum uncertainty packet, the figure shows the numerical values 
for position $k=96$, on a lattice of $2*96$ sites. 
  }\label{fig:unc}
 \end{center}
\end{figure}

Similarly, we find that 
\begin{equation}
x^k_2(t)= ([(M^{-1})^T_{total}] ^t)^k_j x^j_2(0)
\end{equation}
So that 
\begin{equation}
\Delta^2 x^k_1(t) = \left[((M^{-1}_{total}) ^t)^T((M^{-1})_{total} ^t)\right]_{kk}\Delta^2 x^k_1(0)
\end{equation}
 When we plot these, in figure \ref{fig:unc}, we find that the growth in the entropy is essentially linear (it grows like $\log(\Delta x^k_1\Delta x^k_2)$, for times that are short with respect to the travel time in the system.
This indicates that the system behaves (for a long time) as it were chaotic, with an approximate largest Lyapunov exponent given by $\kappa_+\simeq 0.953$, from a linear fit to the logarithm of the uncertainty. 
If the system has periodic boundary conditions there are non-trivial Lyapunov exponents. The maximum Lyapunov exponent for the $192\times192$ matrix of periodic boundary conditions is $\kappa_{max}=0.962424$. So we see that even  the open boundary conditions
that are not chaotic  give a very reasonable estimate to the largest Lyapunov exponent of the periodic system. This is an indication that the open system is behaving in a way that mimics very closely a chaotic system. 

Given this growth of the local entropy at a site, when we impose entropy saturation for a reasonable value of $N\simeq 500$, we find that the entropy of a site saturates before the linear growth in the entropy turns logarithmic in time. 

Consider now the system where we take a region with $s$ sites and also assume that we start the system in a Gaussian state (the same for all sites). The analysis based on Lyapunov exponents \cite{Asplund:2015osa, Berenstein:2015yxu} suggests that the entanglement  entropy of this region should grow as  
\begin{equation}
\frac{\Delta S}{\Delta t} \simeq 2 \lambda_+ s
\end{equation}
Consider now also the bound on the entropy growth that follows from \eqref{eq:qbound} 
\begin{equation}
\frac{\Delta S}{\Delta t} \leq 2 (2 \log (N))
\end{equation}
where we count twice because the interval of $s$ sites will have two sources of entanglement: one on each end of the interval.

These are in contradiction with each other if
\begin{equation}
2 \lambda_+ s \geq 2 (2 \log (N)),
\end{equation} 
or equivalently if
\begin{equation}
s \geq \frac{2 \log (N)}{\lambda_+}
\end{equation}
which indicates that there is a maximum region size where the semiclassical analysis can be valid. The size only grows logarithmically in $N$, and is inversely proportional to the maximum Lyapunov exponent. This indicates that entanglement entropy production will be mostly limited by quantum effects in macroscopic regions for quantum field theories, rather than by the strength of chaos in the system. In a certain sense, the strength of chaos should be thought of as a local property of the entanglement dynamics.

\section{Conclusion}\label{sec:conclusion}

A class of simple lattice models of quantum chaos has been introduced. The models are based on (simple) automorphisms of the algebra of a product (identical) quantum tori.
 A special collection of these can be thought of as a discretized dynamics on rational points of a torus itself. 
 A subset of this class of models was analyzed in more detail. 

The models admit analytic results for analyzing the entanglement entropy evolution for a variety of initial states. 
These models are also simple to simulate if the initial state is factorized between the initial sites.
There are a variety of behaviors observed. First, in some models there can be large classes of states (a whole basis of them)  where the entropy does not evolve in time, even though the state does so. This property is probably peculiar to these models.
A second class of evolution is where the entropy is saturated in one shot and stays high for a while. It is expected that is typical for random initial factorized states. A third class of behavior is where the entropy grows with a rate that is given by a sum of Lyapunov exponents. 
Such behavior is found for semiclassical initial states (states which minimize the uncertainty on all generators of the algebra).	
The evolution of entanglement between regions, when the entangling dynamics is local,  is controlled by the dimension of the Hilbert space on a site $\log(N)$.  When a region is large enough, the semiclassical chaotic dynamics overestimate the entanglement entropy production between (macroscopic) regions, although it can describe accurately the local entangling dynamics at a single site in the lattice. The maximum size region where the semiclassical analysis could possibly be valid scales like $s_{max}\sim \log(N)/\lambda_+$.  One can argue then that  quantum mechanics limits some of the semiclassical effects of chaos and therefore should provide bounds on chaos (this observation has been made previoulsy in \cite{Berenstein:2015yxu}, here I am being more precise about the validity of the semiclassical analysis because the models that have been studied are under better analytic control).

A particularly simple model with $N=2$ can be implemented with quantum CNOT gates and shows very interesting behavior. 
The model acts as a system that can send classical  messages in a lattice in one direction (this is true for any $N$). This tends to produce generalized GHZ states and the eigenvalues for the density matrix for the initial quantum message  are given exactly by the probabilities of the initial state to be in different initial classical message configurations.
This dynamics of entanglement for $N=2$ is controlled by the Siperpinsky triangle fractal and the dynamics shows other self-similar behavior. Once the quantum system is analyzed in a dual (discrete Fourier transform) basis, one can show that in this particular model the entanglement between subregions can be stopped if the 
receiving end of the message is prepared in a different initial state (a product of Hadamard states for the receiving q-bits). The system can thus have all entanglement evolution dynamics restricted to a small interval.

It will be very interesting to analyze this model and similar models for more general classes of initial states (these might Need to be simulated numerically). In particular, the model analyzed here does not seem to naturally admit a notion of temperature, so there is very little one can say about the quantum chaos bound \cite{Maldacena:2015waa}, which was one of the initial motivations to study these models. Maybe some other models will offer more insight in this direction.

The CNOT model can also be implemented in a quantum  computer with current technology. A question that has not been analyzed in this paper, but that it is very interesting,  is how the dynamics gets distorted when the quantum gates are not fully efficient. It is possible that this model and similar models can be used to benchmark quantum computers. In particular, the fact that one can stop entanglement between regions for judiciously chosen initial states in one half (Hadamard product states) describes  is a purely quantum effect, as in the classical dynamics all messages reach their destination. At the same time, the CNOT dynamics discussed here generates states with tractable multi-partite entanglement. Both of these reasons make it interesting to program such evolution on a quantum device.   

The subset of models with discrete dynamics of a rational lattice, which resembles a quantum version of acellular automaton,  is the subset of models where the $Q$ generators of the quantum tori algebra are functions of only $Q$, and the $P$ generators are only functions of $P$.
 More general automorphisms that mix $P,Q$ are possible, but they require fixing some (normal ordering) phases of the automorphism more precisely. These more general models are more interesting, as they do not have a subalgebra of the $Q$ preserved. Thus they are more generic and they are not expected to have a basis of states where no entropy is generated. For such systems it is still possible to compute efficiently various observables with an initial product state, because the automorphism of the algebra still sends monomials to monomials. It is very  interesting to 
explore such models further.

\acknowledgments

The author would like to thank E. Berkowitz, T. Brun,  X. Dong, A. Garcia Garcia, M. Srednicki, D. Weld  for discussions. Work  supported in part by the department of Energy under grant {DE-SC} 0011702.

\end{document}